# First principles molecular dynamics study of filled ice hydrogen hydrate

Jingyun Zhang[1], Jer-Lai Kuo[2*], Toshiaki Iitaka[3*]

[1]School of Physical & Mathematical Sciences, Nanyang Technological University, Singapore

[2]Institute of Atomic and Molecular Sciences, Academia Sinica, Taiwan

[3] Computational Astrophysics Laboratory, RIKEN, Japan

## Abstract

We investigated structural changes, phase diagram, and vibrational properties of hydrogen hydrate in filled-ice phase $C_2$ by using first principles molecular dynamics simulation. It was found that the experimentally reported '*cubic*' structure is unstable at low temperature and/or high pressure. The '*cubic*' structure reflects the symmetry at high (room) temperature where the hydrogen bond network is disordered and the hydrogen molecules are orientationally disordered due to thermal rotation. In this sense, the '*cubic*' symmetry would definitely be lowered at low temperature where the hydrogen bond network and the hydrogen molecules are expected to be ordered. At room temperature and below 30 GPa, it is the thermal effects that play an essential role in stabilizing the structure in '*cubic*' symmetry. Above 60 GPa, the hydrogen bonds in the framework would be symmetrized and the hydrogen bond order-disorder transition would disappear. These results also suggest the phase behavior of other filled-ice hydrates. In the case of rare gas hydrate, there would be no guest molecues rotation-nonrotation transition since the guest molecules keep their spherical symmetry at any temperature. On the contrary methane hydrate MH-III would show complex transitions due to the lower symmetry of the guest molecule. These results would encourage further experimental studies, especially NMR spectroscopy and neutron scattering, on the phases of filled-ice hydrates at high pressures and/or low temperatures.

*To whom corresponding should be addressed.

E-mail: jlkuo@pub.iams.sinica.edu.tw (J.-L.K.), tiitaka@riken.jp (T.I.).

# I. Introduction

Intensive searches for alternative energy resources have been motivated due to the increasing fossil energy consumption and its related global environmental concerns. Hydrates of natural gases such as methane hydrate, $CH_4$-$H_2O$[1,2] (the most abundant natural form of clathrate hydrate) prevailing in deep-sea sediments and permafrost have received active attention and been proposed as one of these energy resources[1]. The global reserve of natural gas in the hydrate form is estimated to be significantly larger than that from traditional fossil fuels and will become a valuable future energy resource[2]. More recently, hydrogen hydrate is of great interest because it facilitates environmentally clean (water is the only by-product) and highly efficient energy conversion. From astronomical point of view, water is known to be a major constituent of Uranus and Neptune.[3-6] On the other side, hydrogen is the most abundant element in the universe, making up 75% of normal matter by mass and over 90% by number of atoms.[7] Water and hydrogen under high pressure may form hydrogen hydrate [8-11]. Therefore, study of water and hydrogen at high pressure has wide implications in astrophysics and knowing their physical properties is also important for understanding the structure and the formation history of these planets. The behavior of hydrates under pressure can also provide valuable information on water–water interactions and interactions of water with a wide range of guest molecules. Furthermore, studying $H_2O$ and $H_2$ mixtures may provide insight into the nature of hydrogen-rich atmosphere in the large-body interstellar ice embryos postulated to exist during planet formation.[12]

At present, three forms of hydrogen hydrates are known to exist stably in different conditions. The first one is a clathrate hydrate, sII structure which was synthesized by compressing a mixture of $H_2$ and $H_2O$ to pressures of 180 to 220 MPa at 300 K.[13-17] and the other two are filled-ice type compounds, $C_1$ and $C_2$, which were synthesized above 0.8 and 2.4 GPa at room temperature, respectively.[8-11] The filled-ice hydrate $C_2$ is composed of hydrogen molecules sitting in the voids of the ice-Ic framework[8] as schematically displayed in Fig.1. The molecular ratios of hydrogen to water are 1:6 and 1:1 for $C_1$ and $C_2$, respectively. In particular, pressure-induced transitions of phase $C_2$ were investigated using X-ray diffraction and Raman spectroscopy.[10, 11] In the low pressure region, $C_2$ hydrogen hydrate remains a cubic crystal structure and structural transitions were observed at approximately 35-40 and 55-60 GPa and the high-pressure phase survived up to at least 80.3 GPa.[11] However, the structures of these two high-pressure phases have not been well refined. There are enormous investigations on hydrogen bond order-disorder transition in ice, like ice Ih/XI[18-24], V/XIII, VI/XV, VII/VIII[25-29], XII/XIV transitions. In this paper, we refine the phase diagram of $C_2$ structure into four phases depending on the ordering of hydrogen bonds and on the rotational state of hydrogen molecules.

## II. Computation methods

The molecular dynamics (MD) simulations at finite temperature and geometry optimization at zero temperature were carried out by means of density functional theory (DFT) calculations using a pseudopotential plane wave approach implemented in Quantum-Espresso Package [18-20] with ultrasoft pseudopotential using the functional of BLYP (from the name Becke for the exchange part and Lee, Yang and Parr for the correlation part)[21]. These simulations are investigated with 8 water molecules and 8 hydrogen molecules at different temperatures (30K, 100K, and 300 K) and in the pressure range from 3 GPa to 100 GPa. The simulations ran at least 15 ps with time step of 0.5 fs. We used a plane wave basis cutoff energy of 35 Ry and a 2×2×2 Monkhorst-Pack (MP) grid (*k*-mesh) for the electronic Brillouin zone integration. A more strict condition with cutoff energy up to 100 Ry and 9×9×6 MP grid was employed in the phonon calculation using the density functional linear-response method.[22] In this work, the vibrational frequencies were also obtained by Fourier transformation of the trajectories obtained from first principles molecular dynamics (FPMD), in which we can trace pressure dependence of the stretching frequencies. We have broadened the theoretical spectra with a Lorentzian of full width at half maximum FWHM = 30 cm$^{-1}$. Another advantage of the method is that anharmonic effects are automatically included in the computed frequencies.

## III. Results and Discussions

### A. Zero temperature results

In order to investigate the structure of hydrogen hydrate at zero temperature, we employed all possible ice-rule-allowed configurations enumeration scheme[23, 24] of hydrogen bond network to construct the framework of ice Ic, which has been successfully applied to ice Ih[25], ice VII/VIII[26, 27], and ice VI[28] in our previous works. Analyzing the symmetry of the all the configurations in the unit cell of ice Ic, there are four class of space groups: $P4_3$, $P4_12_12$, $Pna2_1$ and $I4_1md$. Then, the initial $C_2$ structures were built by putting $H_2$ molecules at the interstitial sites of the ice Ic. We followed two ways to add $H_2$ molecules with different orientations, one is all parallel to each other and the other is orientationally disordered. Therefore, we have 8 $H_2$-$H_2O$ initial structures in total. Full geometry optimizations were carried out for these structures in the pressure range from 5 GPa to 40 GPa and their relative stabilities were shown in Fig. 2, where we could see that arrangements of $H_2$ molecules affect the stability of the structure. Two $H_2$-$H_2O$ structures with space group $P4_12_12$ and $Pna2_1$ are found energetically favorable. Therefore, these two structures are the possible candidate structures of $H_2$-$H_2O$ at zero temperature from the thermodynamic criteria. We chose the structure of $P4_12_12$ with $H_2$ molecules aligned to c-axis for the following phonon dispersion and equation of state (EOS) calculations. The phonon dispersion calculated by density functional linear-response theory show no imaginary frequency between 3 GPa and 60 GPa, indicating the dynamical stability of the structure at zero temperature. It is essential to be noted that the previous reported cubic structure with Fd-3m space group[8] could be applied to the positions of oxygen atoms only at low temperature where the hydrogen bonds and/or hydrogen molecules are ordered. Thus including the hydrogen atoms of water molecules and guest hydrogen molecules would definitely change the symmetry at low temperature and this also explains why we could observe the tetragonal symmetry at low temperature as discussed above. The calculated EOS is compared with the experimental data[11] as shown in Fig. 3. The volume thermal expansion ratio is up to 3% which is in a reasonable range compared to other gas hydrates.[29-31]

## B. Finite temperature effects (300 K)

At high temperature, it is expected that the hydrogen bonds are disordered and the hydrogen molecules are rotating. As discussed above, $H_2$-$H_2O$ structures with space group $P4_12_12$ and $Pna2_1$ are energetically stable and have similar enthalpy in the studied pressure range. These two structures responded similarly under compression in the MD simulations. For simplicity, here we only show the MD results from the ice-Ic framework of space group $P4_12_12$ and $H_2$ molecules aligned to c-axis as the initial structure. At room temperature, the equilibrium structure at low pressure (P< 30 GPa) forms '*cubic*' unit cell, in which the ice-Ic framework remained in the $P4_12_12$ space group symmetry within the simulation time, while $H_2$ molecules were rotating fast at the lattice site. The agreement between theoretical EOS at 300 K and the experimental data is very good below 30 GPa (Fig. 3). This finding demonstrates that the thermal rotation of $H_2$ molecule plays an essential role in stabilizing the structure to appear as *cubic* below 30 GPa, hence, highlights the importance of temperature effects in this system. The discrepancy between the experimental and theoretical volumes above 30 GPa might be attributed to the fact that experimental volume value was calculated assuming the structure remains '*cubic*' symmetry. In fact, the pressure-induced transition from *cubic* to tetragonal symmetry is evidenced by a new diffraction line resulting from the splitting of (220) around 40.5 GPa[11] and the same splitting was also reported around 30-35 GPa in another study[9].

The evolution of lattice constants under pressure and different temperature (0K, 30 K, 100 K and 300 K) was investigated by FPMD (See Fig. 4(b)-(d)). The sudden turn of the lattice constant in *c* direction shown in Fig. 4(a) is caused by the change of compression mechanism. At low pressure, the $H_2$ molecules have different orientations and the compression is almost isotropic. When the applied pressure increases up to 40 GPa, the $H_2$ molecules become ordered and point to one direction which make this direction hard to compress. There is an obvious trend that the temperature stabilizes the '*cubic*' structure from Fig. 4 (b) to Fig. 4 (d). It is easy to see that the structure appears as '*cubic*' ($a = b = c$) at 300K and below 35 GPa and transits into a tetragonal structure ($a = b < c$) at higher pressure (Fig. 4(d)). The transition starts around 40 GPa evidencing by non-identical lattice constants and finally splits around 60 GPa. We double-checked this phenomenon by using larger unit cell consisting of 16 $H_2O$ and 16 $H_2$ molecules and found the same behaviors as discussed above.

Then, we took one more theoretical step towards mapping the phase diagram of the filled-ice Ic hydrogen hydrate as illustrated in Fig. 5. Considering the order-disorder states of the hydrogen bond network and the rotation-nonrotation states of the $H_2$ molecules, there are four combinations. The rotation-nonrotation phase boundary (red line in Fig. 5) was determined by viewing the movie of the MD trajectory at the pressure-temperature points shown in Fig. 5. Note that the rotation of hydrogen molecules will not be perfectly isotropic just above the rotation-nonrotation phase boundary and the crystal symmetry will become cubic at somewhat higher temperature where the molecular rotation becomes isotropic. The order-disorder phase boundary (blue line in Fig. 5) was determined by considering the residual configuration entropy of the ice-Ic framework. It is well known that Pauling[32] deduced the residual configuration entropy of ice with completely disordered hydrogen bond network to be $k_b\ln(3/2) \approx 3.37$ J/mol/K from a pure theoretical estimation.. And this gains remarkably agreement with the later experiment done with ice $I_h$.[33] According to the third law of thermodynamics, there should be no configuration entropy of hydrogen ordered ice. Therefore, we could take the Pauling entropy to estimate the change of entropy at the order-disorder transition. Then considering the formula $\Delta G=TS=k_b T\ln(3/2)$[34], the order-disorder transition is expected around 29 K which is similar to the ice $I_h$-XI transition (around 72 K)[35]. Since the configuration entropy would not depend on the pressure, this phase boundary line is expected to be parallel to the pressure axis. From the experimental point of view, neutron diffraction measurements might be an effective tool to determine the order-disorder boundary, which showed great success in the ice systems, i.e. ice VIII[36-38], ice XI[35, 39-41], ice XIII and ice XIV[42], and ice XV[43]. There are several possible methods for determining the boundary between the hydrogen rotating phase and non-rotating phase. Inelastic neutron scattering was used to study the rotation of guest molecules in methane hydrate[44]. Quasielastic neutron scattering was used to investigate the hydrogen dynamics in crystalline calcium borohydride such as rotations, librations, and vibrations[45]. Proton NMR[46-49] and Raman spectroscopy[10, 11] can detect the rotation of $H_2$ molecules in clathrates. In particular, the diamond anvil cell NMR [49] revealed the molecular rotation and diffusion of $H_2$ molecules in the filled-ice hydrogen hydrate at pressures up to 3.6 GPa.

Vibrational spectroscopy such as Raman spectroscopy and IR absorption spectroscopy is one of the useful materials characterization tools. It can yield information about the form in which hydrogen is present in the material by theoretical assignment of the observed peaks to atomic motion. The vibron (intramolecular vibration mode) of hydrogen molecules is expected to reflect the interaction between hydrogen molecules and the surrounding water molecules and also may relate to the hydrate stabilities in the condition when the hydrogen molecules escape from the water lattice[11]. In our simulation, the vibrational power spectrum or phonon density of state (PDOS) is derived from the Fourier transform of the velocity autocorrelation function[50-52] of the MD trajectories. In Fig. 6, the calculated PDOS at 300 K is shown together with the O-H vibration frequencies calculated from density functional linear-response theory at 0 K and the experimental vibron frequency[11] under compression. There is a good agreement in frequency peak positions between the experimental vibron frequency and the peaks in the PDOS. The blue-shift of the vibron frequency is consistent with the behavior of pure $H_2$ molecules under pressure. The O-H vibration frequencies calculated from density functional linear-response theory also match well with the PDOS obtained from MD trajectories. The O-H vibration frequency decreases as pressure increases until the hydrogen bonds are symmetrized because the O-H bond strength becomes weaker and its bond length becomes longer due to the increased attraction of the proton by the accepting oxygen. Once the hydrogen bonds are symmetrized, the bond length starts decreasing and the vibron frequency starts increasing as pressure increases.

### C. Hydrogen bond symmetrization

The symmetrization of hydrogen bond in ice is intimately related to the quantum motion of protons and has been one of the major subjects in chemistry and physics of ice for over a half century[53-56]. The energy potential felt by the proton in a hydrogen bond between two oxygen atoms can be described as a double-minimum potential. When applying pressure, the potential barrier at the midpoint of the two oxygen atoms is lowered until it disappears and the single well potential forms. Hydrogen-bonded protons initially located at asymmetric positions of the O-O separation will relocate to the symmetric midpoint, which transforms the system to the symmetric phase. Therefore, it has some features of displacive-type phase transitions, namely, soft-mode behavior of the proton related vibrations. The effects of the symmetrization on the vibron frequency and the radial distribution function are studied from the MD trajectories. In Fig. 6, the calculated O-H stretching vibration in $C_2$ hydrogen hydrate (blue squares) softens as a function of pressure and hardens above around 60 GPa. We deduced that the hydrogen bond symmetrized around 60 GPa, which is somewhat higher than the experimental value of 40 GPa[9, 11]. In Fig. 7 the atomic radial distribution function under pressure are shown. The effect of pressure is to shorten O-O distance and to lengthen the O-H bond length. We could see the O-O distance shorten under pressure as shown in g(OO) plot. The first peak position (~ 1 Å) in g(OH) indicates the covalent O-H bond length and the second peak denotes the O.....H hydrogenbond length while the broad band afterwards represents distance from oxygen to the guest $H_2$ molecules. The second peak was observed to move near to first peak and incorporate together around 60 GPa, indicating the hydrogen atoms locate at the middle point between two neighboring oxygen atoms. To further investigate the thermal hopping behavior of the hydrogen atoms, we plotted the probability distribution of hydrogen atom along a hydrogen bond at room temperature in Fig. 8. Benoit *el al*[57] proposed a three stage scenario theory for illustrating the hydrogen bond symmetrization of pure ice with increasing pressure in which under low pressure the ice is a molecular state with normal hydrogen bond, then under medium pressure hydrogen atoms start to jump between two potential minimum (tunneling or ionization), finally under high pressure hydrogen atoms move to the midpoint between two neighboring oxygen atoms (symmetrization). From Fig. 8, the probability distribution of d(OH)/d(OO) clearly shows that the three-stage scenario is also valid for hydrogen hydrate as it was shown for the filled ice $I_h$ structure of methane hydrate[58-60].

Moreover, we would like to comment on the relation between hydrogen bond symmetrization of $C_2$ hydrogen hydrate and that of ice VII. The O-O distance and the ratio of d(OH)/d(OO) in $C_2$ hydrogen hydrate and in ice VII calculated at zero temperature were shown in Fig. 9 together with experimental data. The calculated O-O distance in Fig. 9(a) coincides well with experimental results below 30 GPa while large difference was observed in higher pressure region. A possible explanation would be the experimental results were obtained by assuming '*cubic*' symmetry in all the pressure range and calculating the O-O distance directly by the lattice parameter *a* as: $d_{O-O} = \frac{1}{4}\sqrt{3}a$ and this would cause error as also in the EOS. The predicted distances between two neighboring oxygen atoms ($d_{O-O}$) of 2.44, 2.36, and 2.27 Å are the onset of proton tunneling and unimodal distribution, and single well potential[57, 61] as indicated by horizontal bars in Fig. 9(a) and Fig. 9(b). We could see that the hydrogen bond symmetrization (dOH/dOO = 0.5) took place around 60 GPa and 120 GPa for $C_2$ hydrogen hydrate and ice VII, respectively, deduced from Fig 9(c) and Fig. 9(d). The higher H-bond symmetrization pressure compared to the experimental value (40 GPa for hydrogen hydrate[11] and 62 GPa for ice VII[62]) can be greatly alleviated by taking quantum effects of hydrogen motion into account.[57] In comparison with $d_{O-O}$ calculations in ice VII, we found that the pressure dependence of $d_{O-O}$ of $C_2$ hydrogen hydrate is significantly larger than ice VII and this dependence could be mapped onto that of ice VII by doubling the pressure which was also indicated in an experimental report[9] and attributed to the fact that $C_2$ hydrogen hydrate consists of only one of the two cubic sublattices in ice VII. From this point of view, the properties of $C_2$ hydrogen hydrate at a given pressure are a good model for $H_2O$ at approximately twice that pressure.

## IV. Summary

In this work, we investigated the structural changes, phase diagram, and vibrational properties of hydrogen hydrate in filled-ice phase $C_2$ by using first principles molecular dynamics simulation. It was found that the experimentally reported '*cubic*' structure with space group Fd-3m [8] is unstable at low temperature and/or high pressure: the '*cubic*' structure reflects the symmetry at high (room) temperature where the hydrogen bond network is disordered and the hydrogen molecules are orientationally disordered by thermal rotation. In this sense, the '*cubic*' symmetry would definitely be lowered at low temperature where the hydrogen bond network and the hydrogen molecules are expected to be ordered. Actually, two $H_2$-$H_2O$ structures with space group $P4_12_12$ and $Pna2_1$ are found comparatively energetically favorable at 0 K and between 5 GPa and 40 GPa. At room temperature and below 30 GPa, it is the thermal effects that play an essential role in stabilizing the structure in '*cubic*' symmetry. We also observed a phase transition to an unknown new phase around 40 GPa at room temperature in agreement with recent experimental finding [11]. Above 60 GPa, the hydrogen bonds in the framework would be symmetrized and the hydrogen bond order-disorder transition would disappear. These results also suggest the phase behavior of other filled-ice hydrates. In the case of rare gas hydrates [63, 64], there would be no guest molecules rotation-nonrotation transition since the guest molecules keep their spherical symmetry at any temperature. On the contrary methane hydrate MH-III [58-60] would show complex transitions due to the lower symmetry of the guest molecule.

In summary, we estimated the phase diagram of the hydrogen hydrate $C_2$ structure with four phases (hydrogen bond ordered-disorderd, $H_2$ rototaion-nonrotation) at a first approximation. These results would encourage further experimental studies, especially NMR spectroscopy and neutron scattering, on the phases of filled-ice hydrates at high pressures and/or low temperatures.

## Acknowledgment

The authors thank Dr. Quoc Chinh Nguyen for his help in numerical calculation. All the calculations were performed by using the RIKEN Integrated Cluster of Clusters (RICC) facility. This work is financially supported by Nanyang Technological University, Academia Sinica, the National Science Council (NSC98-2113-M-001-029-MY3) of Taiwan and KAKENHI (No. 20103001, No. 20103005) from MEXT of Japan.

# Figure Captions

Fig. 1 (Color online) Four schematic models of structure. (a) space group $P4_12_12$ ; (b) space group $Pna2_1$ ; (c) space group $I4_1/amd$ . H atoms of $H_2O$ have half occupation indicating the hydrogen bond disordered network; (d) space group $Fd-3m$. H atoms of $H_2O$ have half occupation. The rotating guest $H_2$ molecules are represented by a light blue sphere; The red balls represent oxygen atoms in water molecules, the yellow balls indicate hydrogen atoms in water molecules, the grey balls indicate the hydrogen atoms of the hydrogen molecules.

Fig. 2 (Color online) Realtive enthalpies of 8 $H_2$-$H_2O$ candidate structures as a function of pressure.

Fig. 3 Comparison of calculated equation of states of structure with $P4_12_12$ space group while $H_2$ aligned at 0 K and 300 K (solid line and solid squares with error bar, respectively) with experimental data (open squares with error bar).

Fig. 4 Lattice constant evolution at 0K , 30K, 100K, and 300 K as a function of pressure in which *a*, *b*, *c* was indicated as empty squares, triangles, and stars, respectively.

Fig. 5 (Color online) Phase diagram of $H_2$-$H_2O$'s four possible states and all the (P,T) conditions in our MD simulation shown in solid squares.

Fig. 6 (Color online) Comparison of simulated phonon density of states (solid line) from MD trajectory at 300 K with the frequency (solid blue squares) calculated with density functional linear response theory and the experimental data in [11] (red symbols).

Fig. 7 Radial distribution function g(OO) and g(OH) as a function of pressure of hydrogen hydrate at 300K.

Fig. 8 Distribution of hydrogen atom along a hydrogen bond of $C_2$ hydrogen hydrate at 300 K and 3, 30, 60 GPa, respectively.

Fig. 9 Calculated oxygen-oxygen distance and the ratio d(OH)/d(OO) in $C_2$ hydrogen hydrate and ice VII at 0 K.

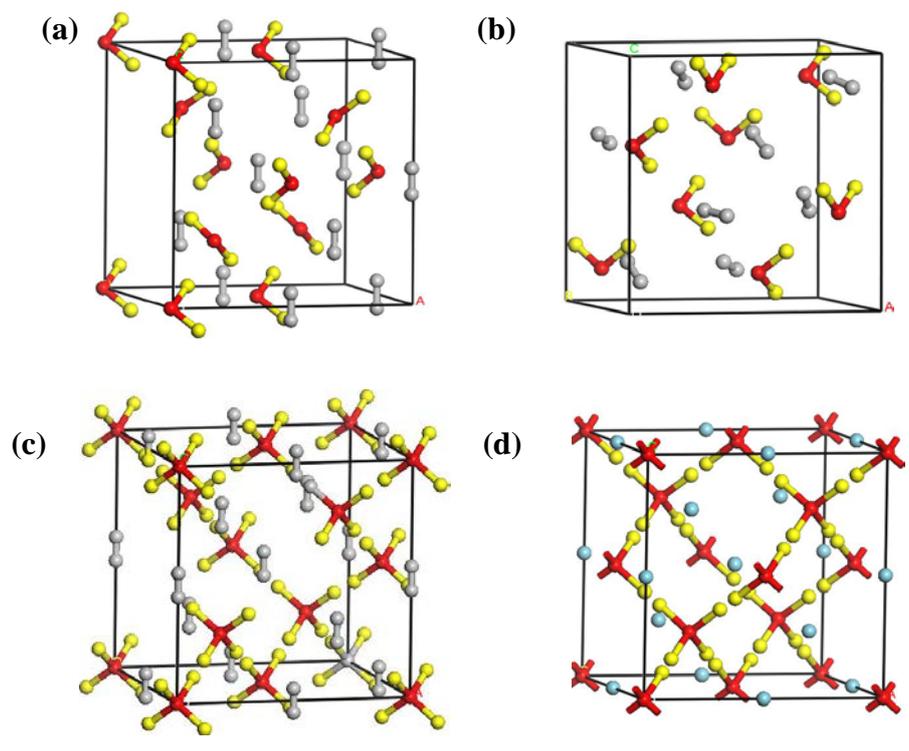

**FIG. 1**

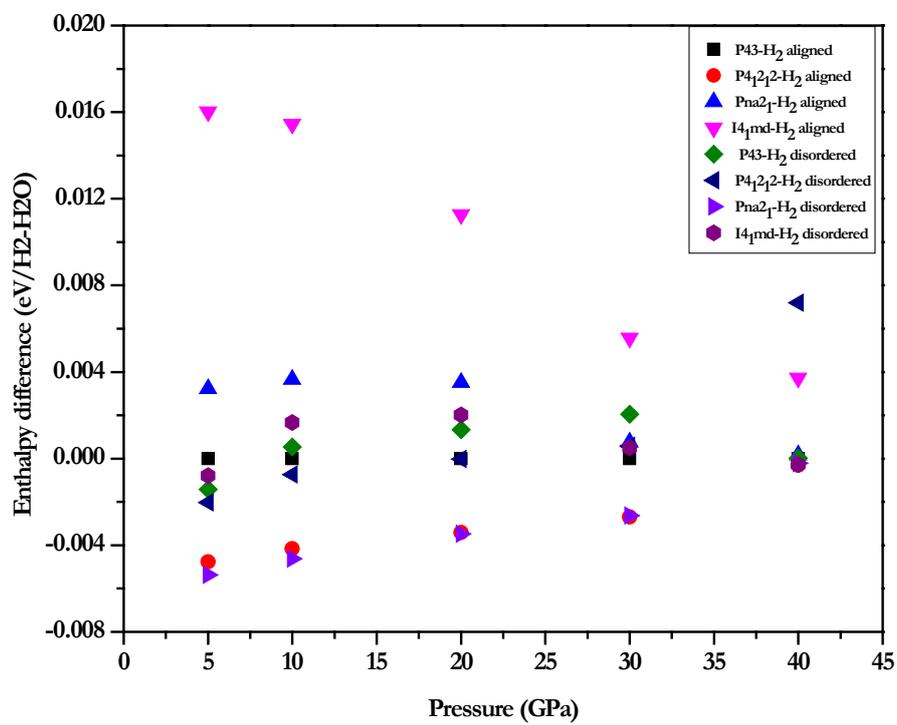

**FIG. 2**

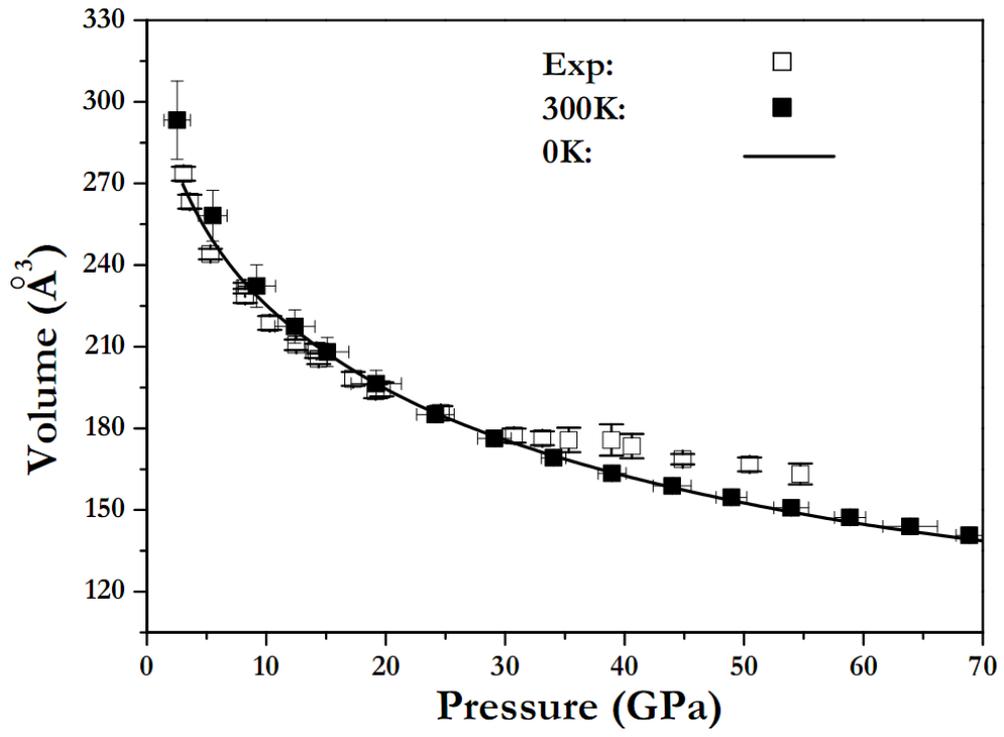

**FIG. 3**

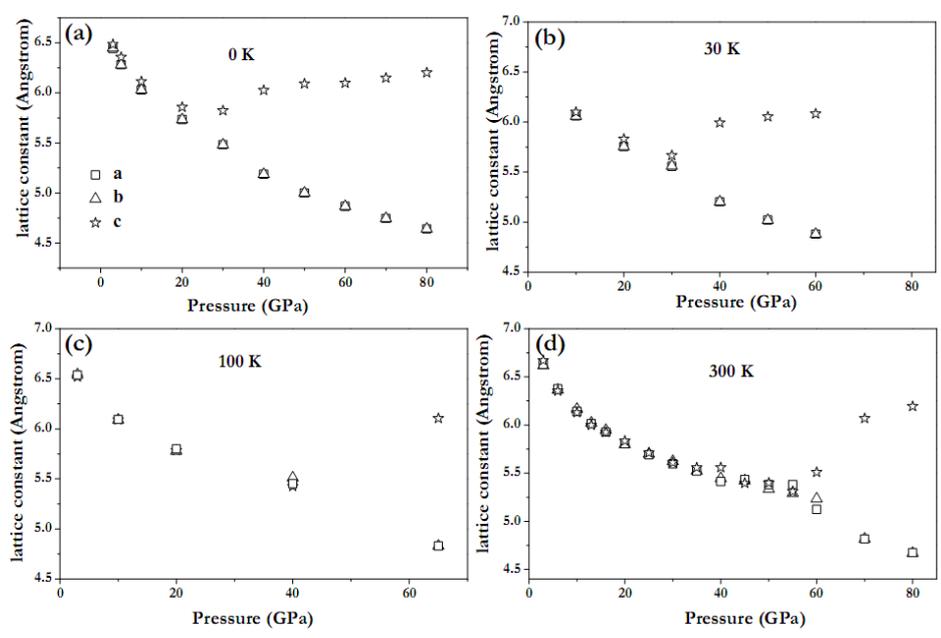

**FIG. 4**

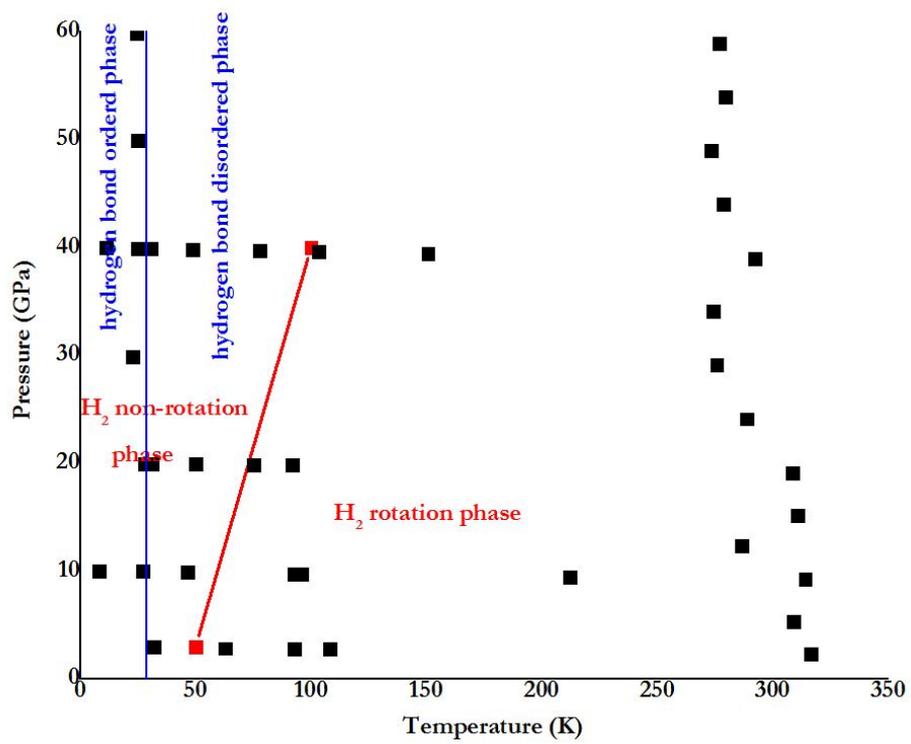

**FIG. 5**

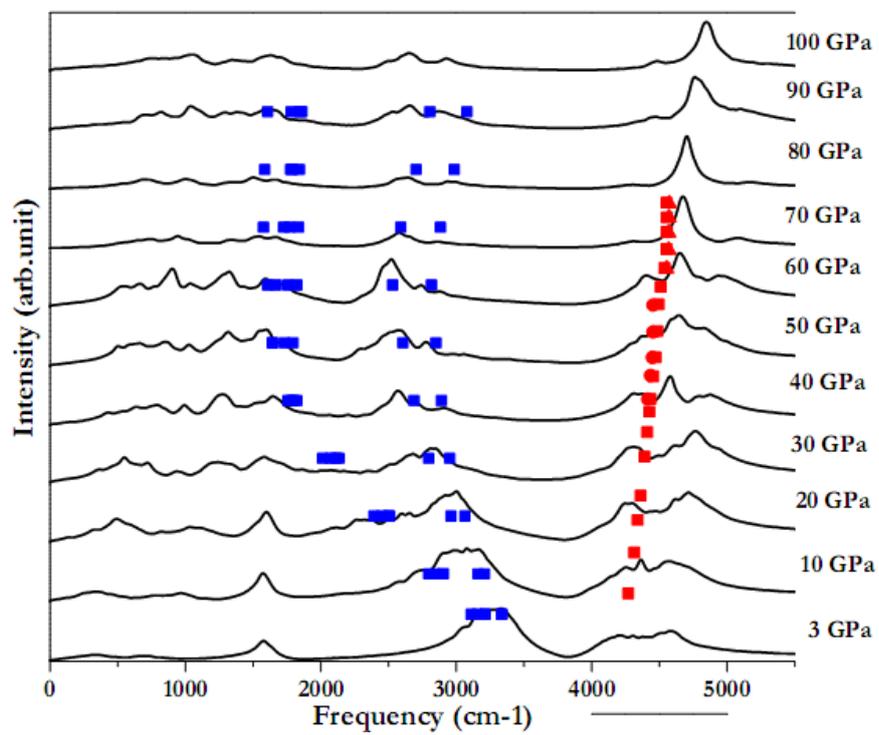

**FIG. 6**

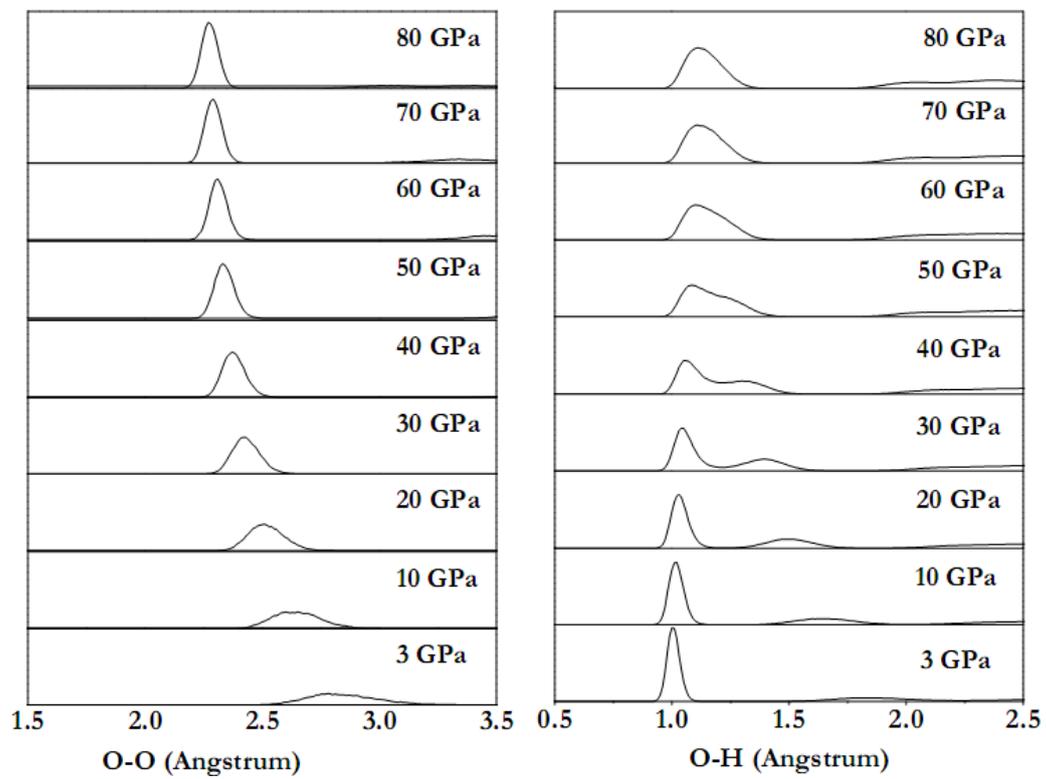

**FIG. 7**

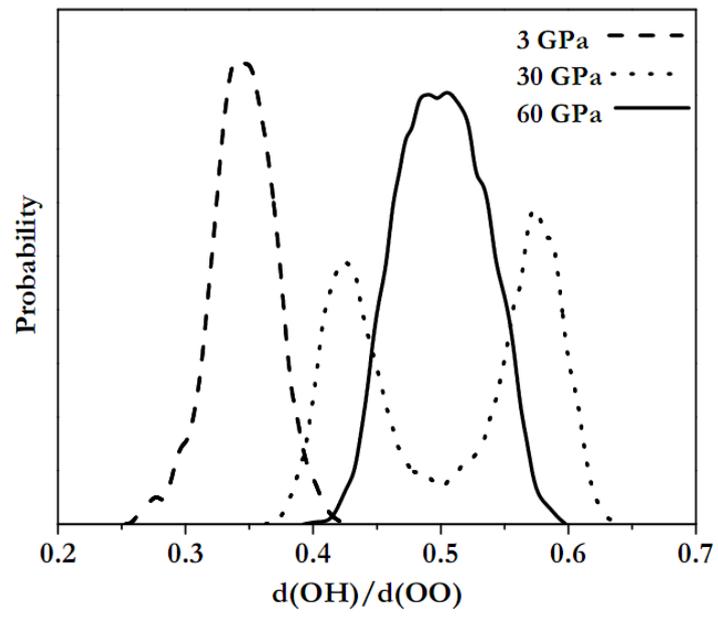

**FIG. 8**

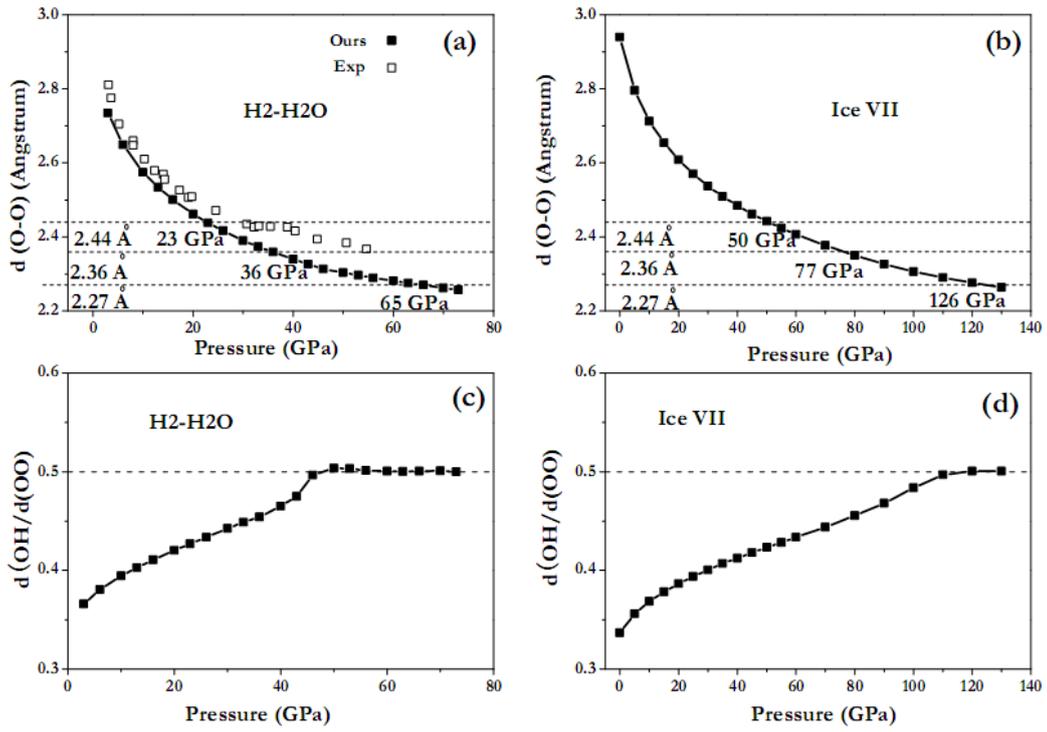

**FIG. 9**